\documentclass[aoas,MSNbibl,nameyear,seceqn,dvips]{arximspdf}
\usepackage{dcolumn}
\usepackage{graphicx}

\doi{10.1214/12-AOAS581} 
\volume{6}
\issue{4}
\pubyear{2012}
\firstpage{1949}
\lastpage{1970}

\makeatletter
\newcolumntype{d}[1]{D{.}{.}{#1}}
\newcommand{\rrvert}{\vert}
\newcommand{\llvert}{\vert}
\newcommand{\pr}{\operatorname{pr}}
\newcommand{\tr}{^\mathrm{T}}

\newcommand{\ftheta}{\bolds{\theta}}
\newcommand{\fmu}{\bolds{\mu}}
\newcommand{\fbeta}{\bolds{\beta}}
\newcommand{\fu}{\mathbf{u}}
\makeatother

\begin{document}
\begin{frontmatter}

\title{The ranking lasso and its application to sport tournaments\thanksref{T1}}
\thankstext{T1}{Supported by PRIN 2008 grant.}
\runtitle{The ranking lasso}

\begin{aug}
\author[A]{\fnms{Guido} \snm{Masarotto}\ead[label=e1]{guido.masarotto@unipd.it}}
\and
\author[B]{\fnms{Cristiano} \snm{Varin}\corref{}\ead[label=e2]{sammy@unive.it}}
\runauthor{G. Masarotto and C. Varin}
\affiliation{Universit\`a di Padova and Universit\`a Ca' Foscari Venezia}
\address[A]{Department of Statistical Sciences\\
Universit\`a di Padova\\
Via Cesare Battisti, 241 \\
I-35121 Padova\\
Italy\\
\printead{e1}} 
\address[B]{Department of Environmental Sciences\\
Informatics and Statistics\\
Universit\`a Ca' Foscari Venezia\\
San Giobbe Cannaregio, 873 \\
I-30121 Venice\\
Italy\\
\printead{e2}}
\end{aug}

\received{\smonth{11} \syear{2011}}
\revised{\smonth{6} \syear{2012}}

%
\begin{abstract}
Ranking a vector of alternatives on the basis of a series of paired
comparisons is a relevant topic in many instances. A popular example
is ranking contestants in sport tournaments. To this purpose, paired
comparison models such as the Bradley--Terry model are often used.
This paper suggests fitting paired comparison models with a lasso-type procedure
that forces contestants with similar abilities
to be classified into the same group. Benefits of the proposed method
are easier
interpretation of rankings and a significant improvement of the quality
of predictions with respect to the standard maximum likelihood fitting.
Numerical aspects of the
proposed method are discussed in detail. The methodology is
illustrated through ranking of the teams of the National Football
League 2010--2011 and the American College Hockey Men's Division I 2009--2010.
\end{abstract}

%

\end{frontmatter}

\section{Introduction}\label{sectintroduction}

Paired comparison data arise when a series of alternatives is compared
in pairs, typically with the aim of producing a ranking or identifying
predictors of future comparisons. Since the pioneering work of
\citet{thurstone1927}, a considerable amount of literature has been
published on modeling paired comparison data, especially in the wide
field of social sciences. See the recent reviews by
\citet{bockenholt2006} and \citet{cattelan2011}.

Paired comparison data are also the norm in sport tournaments, where
teams play matches against each other. When the round-robin
(all-play-all) tournaments cannot be scheduled as in North-American
major league sports, rankings based on the
records of victories-ties-defeats are questionable because teams may
have a sensible advantage or disadvantage from the skill level of the
other teams within the same division and within the same
conference. This tournament design issue motivated a variety of ranking
procedures either based on scientific methods or on subjective
evaluations, such as votes\vadjust{\goodbreak} from pools of experts. A case that also yields
much interest within the statistical community is the
identification of a champion of the US college football; see the paper
by \citet{stern2004} and its discussion.

Rankings derived from paired comparison models have been proposed for
several sports, such as American football [\citet
{glickman1999,mease2003}], association football [\citet{fahrmeir1994,held2000}],
basketball [\citeauthor{held2000}], chess [\citet{joe1990,glickman1999}]
and tennis [\citeauthor{glickman1999} (\citeyear{glickman1999,glickman1999,glickman2001})]. In these papers, authors suggest variants
of the basic paired comparison models to provide sensible rankings or
improve predictions of future results.

In this paper we argue in favor of
rankings constructed so that teams with similar abilities are
classified into the same group. In order to obtain rankings in groups,
we propose to fit a
paired comparison model with a lasso-type penalty
[\citet{tibshirani1996}]. To the best of our knowledge, this is the
first time that a lasso-type penalty is used in conjunction with a
paired comparison model for the purpose of ranking. Benefits of the
proposed ranking in groups procedure are twofold. First, interpretation
of ranking is simplified by grouping, especially when the number of
teams is not small and there are teams with similar ability. Then, the
shrinkage of the lasso procedure significantly improves the quality of
predictions with respect to the standard maximum likelihood
fitting. The proposed methodology is illustrated through analysis of
the regular season of the National Football League
(NFL) 2010--2011 and of the NCAA American College Hockey Men's Division
I 2009--2010.

The paper is organized as follow. First, analyses of NFL data with
standard paired comparison models are
presented in Section~\ref{sectBTmodel}. Section~\ref{sectranking}
presents our lasso-type method for ranking in
groups. The application to the NFL tournament is given in Section \ref
{sectNFL}. Section~\ref{sectties} describes the extension to sport with
possible ties and illustrates it with the analysis of the NCAA hockey
tournament.

\section{Bradley--Terry rankings}\label{sectBTmodel}

Although the methodology discussed in this paper is of potential
interest for any situation where $k$ treatments are compared pairwise,
thereafter sport terminology is used because of our specific
application. Consider a tournament involving $k$ teams and denote by
$Y_{ijr}$ the random variable for the outcome of the $r$th match
between team $i$ and team $j$. We start by considering only
sports whose rules do not allow for ties, hence, $Y_{ijr}$ is the
Bernoulli variable
\[
Y_{ijr}=\cases{ 1, &\quad$\mbox{if team $i$ defeats team $j$,}$\vspace
*{2pt}
\cr
0, &\quad \mbox{if team $i$ is defeated by team $j$,} }
\]
with $r=1, \ldots, n_{ij}$. The total number
of matches is denoted by $n=\sum_{i< j}^k n_{ij}$. The extension of
the model to handle ties is illustrated in Section~\ref{sectties}.

A popular statistical model for ranking teams in tournaments is the
Bradley--Terry model [\citet{bradley1952}]. This is a logistic\vadjust{\goodbreak}
regression model
%
\begin{equation}
\label{eqnBT} \pr(Y_{ijr}=1)=\frac{\exp(\tau
h_{ijr}+\mu_i-\mu_j)}{1+\exp(\tau h_{ijr}+\mu_i-\mu_j)},
\end{equation}
where $h_{ijr}$ is the home-field indicator for the $r$th game between
teams $i$ and $j$ defined as follows:
\[
h_{ijr}=\cases{ 1, & \quad$\mbox{if the match is played at home of team
$i$,}$\vspace *{2pt}
\cr
0, & \quad$\mbox{if played on neutral field}$,\vspace*{2pt}
\cr
-1, & \quad$\mbox{if played at home of team $j$.}$}
\]
The model parameters are the home-field parameter
$\tau$ and the vector of team abilities $\fmu=(\mu_1, \ldots,
\mu_k)\tr$. Alternatively, one could consider separate home-field
parameters $\tau_i$ for each team. However, as observed by
\citet{mease2003}, this refinement is of little benefit for the
purpose of ranking because then it requires distinct rankings for
teams when playing at home or away.

Model (\ref{eqnBT}) is identified through the pairwise differences
$\mu_i-\mu_j.$ Hence, it is necessary to include one contrast on the
abilities vector, such as $\mu_1=0$ or the sum contrast $\sum_{i=1}^k
\mu_i=0$. We choose the second option since it facilitates
communication to a nontechnical audience.

The inferential target of the analysis is to estimate the abilities
vector and then use this for ranking the $k$ teams. The standard
analysis relies on the maximization of the log-likelihood computed
under the
assumption of the independence among the matches
%
\begin{equation}\qquad\quad
\label{eqnlik} \ell(\fmu, \tau)=\sum_{i<j}^k
\sum_{r=1}^{n_{ij}} y_{ijr} (\tau
h_{ijr}+\mu_i-\mu_j)-\log \bigl\{1+\exp(\tau
h_{ijr}+\mu_i-\mu_j) \bigr\}.
\end{equation}
Maximum likelihood estimation for this Bradley--Terry
model can be performed through standard software for generalized
linear models or using specialized programs as the \texttt{R}
[\citet{R2011}] package \texttt{BradleyTerry2} [\citet{turner2011}].

\begin{table}
\def\arraystrech{0.9}
\caption{NFL regular season 2010--2011. For each team, the table
displays the record and the ability estimated by maximum likelihood
(\texttt{MLE}), by adaptive ranking lasso (\texttt{lasso}) and by
hybrid adaptive ranking lasso/maximum likelihood (\texttt{hybrid}).
Results are
shown with both \texttt{AIC} and \texttt{BIC} model selection. Teams
qualified for playoff are marked by symbol $^\dag$}\label{tabranking}
\vspace*{-3pt}
\begin{tabular*}{\textwidth}{@{\extracolsep{\fill}}lcd{2.2}d{2.2}d{2.2}d{2.2}d{2.2}@{}}
\hline
&&& \multicolumn{2}{c}{\textbf{Lasso}} & \multicolumn{2}{c@{}}{\textbf{Hybrid}} \\[-4pt]
&&& \multicolumn{2}{c}{\hrulefill} & \multicolumn{2}{c@{}}{\hrulefill} \\
\multicolumn{1}{@{}l}{\textbf{Teams}}& \multicolumn{1}{c}{\textbf{Record}} & \multicolumn{1}{c}{\textbf{MLE}} &
\multicolumn{1}{c}{\textbf{AIC}} & \multicolumn{1}{c}{\textbf{BIC}} & \multicolumn{1}{c}{\textbf{AIC}} & \multicolumn{1}{c@{}}{\textbf{BIC}} \\
\hline
New England Patriots$^\dag$ & 14--2 & 2.59 & 1.40 & 1.13 & 2.56 &
2.54 \\
Atlanta Falcons$^\dag$ & 13--3 & 1.82 & 0.76 & 0.53 & 1.78 & 1.73 \\
Baltimore Ravens$^\dag$ & 12--4 & 1.75 & 0.76 & 0.53 & 1.78 & 1.73
\\
Pittsburgh Steelers$^\dag$ & 12--4 & 1.74 & 0.76 & 0.53 & 1.78 &
1.73 \\
New York Jets$^\dag$ & 11--5 & 1.37 & 0.59 & 0.40 & 1.35 & 1.35 \\
Chicago Bears$^\dag$ & 11--5 & 1.00 & 0.28 & 0.10 & 0.91 & 0.87 \\
New Orleans Saints$^\dag$ & 11--5 & 0.93 & 0.28 & 0.10 & 0.91 &
0.87 \\
Green Bay Packers$^\dag$ & 10--6 & 0.91 & 0.28 & 0.10 & 0.91 & 0.87
\\
Tampa Bay Buccaneers & 10--6 & 0.61 & 0.04 & -0.11 & 0.55 & 0.32 \\
Philadelphia Eagles$^\dag$ & 10--6 & 0.49 & 0.04 & -0.11 & 0.55 &
0.32 \\
New York Giants & 10--6 & 0.33 & -0.02 & -0.11 & 0.23 & 0.32 \\
Indianapolis Colts$^\dag$ & 10--6 & 0.20 & -0.02 & -0.11 & 0.23 &
0.32 \\
Miami Dolphins & 7--9 & 0.19 & -0.02 & -0.11 & 0.23 & 0.32 \\
Kansas City Chiefs$^\dag$ & 10--6 & -0.16 & -0.21 & -0.12 & -0.56 &
-0.63 \\
Detroit Lions & 6--10 & -0.21 & -0.21 & -0.12 & -0.56 & -0.63 \\
Minnesota Vikings & 6--10 & -0.28 & -0.21 & -0.12 & -0.56 & -0.63 \\
San Diego Chargers & 9--7 & -0.28 & -0.21 & -0.12 & -0.56 & -0.63 \\
Cleveland Browns & 5--11 & -0.38 & -0.21 & -0.12 & -0.56 & -0.63 \\
Jacksonville Jaguars & 8--8 & -0.39 & -0.21 & -0.12 & -0.56 & -0.63
\\
Oakland Raiders & 8--8 & -0.53 & -0.21 & -0.12 & -0.56 & -0.63 \\
Washington Redskins & 6--10 & -0.56 & -0.21 & -0.12 & -0.56 & -0.63
\\
Dallas Cowboys & 6--10 & -0.58 & -0.21 & -0.12 & -0.56 & -0.63 \\
Buffalo Bills & 4--12 & -0.67 & -0.21 & -0.12 & -0.56 & -0.63 \\
Houston Texans & 6--10 & -0.71 & -0.21 & -0.12 & -0.56 & -0.63 \\
Tennessee Titans & 6--10 & -0.74 & -0.21 & -0.12 & -0.56 & -0.63 \\
Seattle Seahawks$^\dag$ & 7--9 & -0.76 & -0.21 & -0.12 & -0.56 &
-0.63 \\
Cincinnati Bengals & 4--12 & -0.78 & -0.21 & -0.12 & -0.56 & -0.63
\\
St Louis Rams & 7--9 & -0.86 & -0.21 & -0.12 & -0.56 & -0.63 \\
San Francisco 49ers & 6--10 & -1.03 & -0.21 & -0.12 & -0.56 & -0.63
\\
Arizona Cardinals & 5--11 & -1.42 & -0.39 & -0.12 & -1.43 & -0.63 \\
Denver Broncos & 4--12 & -1.54 & -0.39 & -0.12 & -1.43 & -0.63 \\
Carolina Panthers & 2--14 & -2.02 & -0.93 & -0.74 & -1.91 & -1.86 \\
\hline
\end{tabular*}       \vspace*{-3pt}
\end{table}

\subsection{NFL regular season 2010--2011}

The 2010--2011 regular season of the National Football League (NFL)
involves thirty-two teams evenly partitioned into two conferences,
called the American Football Conference (AFC) and the National
Football Conference (NFC). The two conferences are subdivided into
four regional divisions with four teams each. The regular season
consists of $16$ matches per team, scheduled in such a way to
guarantee six matches (three at home and three away) against the other
teams of their own division, six matches (three at home and three
away) against teams of other divisions in their own conference and
four matches (two at home and two away) against teams of the other
conference. The last regular season of NFL thus involved $256$ matches
scheduled into $17$ weeks from September 9, 2010 to January 2,\vadjust{\goodbreak}
2011. Formally, regular season matches could end with a tie, but ties
are very infrequent since the institution of the overtime period in
1974. Indeed, there have been only 17 tie games since 1974, and none
occurred during season 2010--2011. In this season $143$ matches out of
$256$ were won by the home team ($55.8\%$), thus suggesting a slight
home advantage. The second column of Table~\ref{tabranking} reports
the record of victories-losses for each of the $32$ teams during the
regular season.\vadjust{\goodbreak}

We proceed now with maximum likelihood analysis of the Bradley--Terry
model. The maximum likelihood estimate of the home field parameter
is $\hat\tau^{\mathrm{(mle)}}=0.322,$ with a standard error equal to
$0.149$, thus supporting the evidence of a positive effect of playing
on their home field. Maximum likelihood estimates of teams abilities
$\hat
\mu^{\mathrm{(mle)}}_i$, computed under the sum contrast, are reported
in the third column of Table~\ref{tabranking}. The New England Patriots
is the team with the highest estimated ability during the regular
season. This result confirms the top record of the team with~14
victories and two defeats only. In fact, the top seven teams
according to the estimated Bradley--Terry model are also those with
the best records.

The concordance between the ranking of the maximum likelihood
estimates of the abilities and the frequency of victories does not
hold for all the teams. For example, the Miami Dolphins with a record of
seven victories and nine defeats has an estimated ability of $0.189$,
which is sensibly larger than the estimated ability of the Kansas City
Chiefs equal to $-0.158$, although this team has a better record of ten
victories and six defeats. This result is explained by looking more
closely at the results of the matches played by the two teams. In
fact, while Kansas played only teams of similar or lower ability with
alternating results, the Dolphins also played teams with a better record,
in two cases succeeding against the Green Bay Packers and the New York Jets.

At the end of the regular season twelve teams are qualified to the playoff.
The first twelve teams of the Bradley--Terry ranking include
ten of the teams actually qualified to the playoff; see Table~\ref{tabranking}.
The two qualified teams excluded are the Kansas City Chiefs,
which is, however, close to the top 12 since it is ranked at the $14$th
position, and the Seattle Seahawks,
which instead has a very low $26$th position. In place of these
two teams, the Bradley--Terry ranking promotes the Tampa Bay Buccaneers
and the New York
Giants.

\section{The ranking lasso}\label{sectranking}

As anticipated, in this paper we argue in favor of ranking in groups
formed by ``statistically equivalent'' teams.
Ranking in groups is obtained by maximizing
the Bradley--Terry likelihood (\ref{eqnlik}) with a $L_1$ penalty on all
the pairwise
differences of abilities $\mu_i-\mu_j,$
%
\begin{equation}
\label{eqnrls} (\hat{\fmu}_{\lambda}, \hat {\tau}_{\lambda})=\arg\max
\ell(\fmu, \tau) \qquad\mbox{subject to } \sum_{i<j}^k
w_{ij} | \mu_i-\mu_j | \leq s,
\end{equation}
where $w_{ij}$ are pair-specific weights. A particular
choice of the weights is discussed in Section~\ref{sectadapt}. The
standard maximum likelihood solution is obtained for a sufficiently
large value of the bound $s$, while fitting is penalized as $s$
decreases to zero, resulting in groups of team ability parameters that
are estimated to the same value.
Thereafter, the process of solving problem
(\ref{eqnrls}) is termed the \textit{ranking lasso} method. However, the
proposed method
does not merely produce a ranking of the teams but a rating also suitable
for prediction, as illustrated in Section~\ref{sectpredict}. Hence, an
alternative valid name for the proposed method is \textit{rating
lasso}.

The ranking lasso problem is equivalent to the penalized minimization
problem
%
\begin{equation}
\label{eqnrl} (\hat{\fmu}_\lambda, \hat {\tau}_\lambda)=\arg\min
\Biggl\{-\ell(\fmu, \tau) +\lambda\sum_{i<j}^k
w_{ij} | \mu_i-\mu_j | \Biggr\}
\end{equation}
for a certain penalty $\lambda$ that has a one-to-one
relation with the bound $s$. The following reformulation of the
ranking lasso problem as a constrained ordinary lasso problem is
useful for subsequent developments
%
\begin{eqnarray}
\label{eqnrlmin} %
(\hat{\fmu}_\lambda, \hat
\tau_{\lambda},\hat{\ftheta}_\lambda )=\arg \min \Biggl\{-\ell(\fmu,
\tau)+\lambda\sum_{i<j}^k w_{ij}
| \theta_{ij} | \Biggr\}
\nonumber
\\[-8pt]
\\[-8pt]
\eqntext{\mbox{subject to } \theta_{ij}=\mu_i-\mu_j,
1<i<j<k.}
\end{eqnarray}

The penalty used in the ranking lasso is a generalization of the fused
lasso penalty [\citet{tibshirani2005}]. The fused lasso is designed for
problems where the coefficients to be shrunk have some natural order
so that only pairwise differences of successive coefficients need to
be penalized. The lack of order in the ranking lasso implies
substantial computational difficulties. Essentially, the complications
arise because of the one-to-many relationship between the coefficients
of interest $\mu_i$ and the penalized parameters
$\theta_{ij}=\mu_i-\mu_j,$ $i<j$. In Section~\ref{sectsolution} we
supply a convenient numerical approach to compute the solution of the
ranking lasso.

Recently, a certain interest has been paid to linear regression models
for continuous responses with \textit{generalized fused lasso}
penalties, that is, $L_1$ penalties on generic sets of pairwise
differences of parameters. \citet{she2010} investigates the use of
this type of penalty to perform unsupervised clustering in microarray
data analysis. The resulting penalized method has been termed the
clustered lasso. \citet{she2010} provides asymptotic properties of the
clustered lasso and develops an annealing-type algorithm to compute
its solution. \citet{bondell2009} propose a generalized fused lasso
approach for factor selection and level fusion in
ANOVA. \citet{gertheiss2010} use the same penalty to evaluate the
levels of a nominal categorical variable that should be collapsed
together. To this aim, they approximate the lasso solution by
introducing a quadratic penalty method. \citet{guo2010} suggest to
shrink the differences between every pair of cluster centers in
high-dimensional model-based clustering. Optimization is then
performed via an expectation--maximization algorithm where the $L_1$
penalty is substituted by a local quadratic
approximation. \citet{tibshirani2010} develop a path algorithm for
generalized lasso problems with penalty $\lambda|\mathbf{D} \fbeta
|,$ where
$\fbeta$ are regressor coefficients and $\mathbf{D}$ is a matrix not
necessarily
of full rank. Hence, the generalized lasso includes the generalized
fused lasso as a special case. The key idea of \citet{tibshirani2010}
is to overcome numerical difficulties by solving the simpler Lagrange
dual problem.

\subsection{The adaptive ranking lasso}\label{sectadapt}

As noted by various authors [e.g., \citet{fan2001}, \citeauthor{zou2006} (\citeyear{zou2006})], the lasso method
can yield
inconsistent estimates of the nonzero effects because the shrinkage
produced by the $L_1$ penalty is too severe. In terms of the ranking
lasso, this inconsistency means that the bias of the estimators of the
nonzero pairwise differences of abilities does not decrease to zero as
the number of matches raises.

This drawback of lasso can be overcame by employing different
data-dependent penalties in such a way to preserve true large effects.
A first possibility is to substitute the $L_1$ penalty with a
continuous penalty that penalizes large effects less severely. This
idea is implemented in the smoothly clipped absolute deviation (SCAD)
method suggested by \citet{fan2001}. An alternative, which we follow in
this paper, is to weight more the terms of the $L_1$ lasso penalty as
the size of the effect decreases. The adaptive lasso [\citet{zou2006}]
follows this strategy using weights inversely proportional to the
maximum likelihood estimates.

Accordingly, we identify the \textit{adaptive ranking lasso}
method as the solution of~(\ref{eqnrls}) with weights inversely
proportional to the maximum likelihood estimates
%
\begin{equation}
\label{eqnadaweights} w_{ij}=\bigl\llvert \hat\mu_i^{\mathrm{(mle)}}-
\hat\mu_j^{\mathrm{(mle)}} \bigr\rrvert^{-1},
\end{equation}
so as to protect large differences of abilities. The rationale is that
as the sample size raises, then weights given to nonzero pairwise
differences of abilities converge to a finite constant, while the
weights for the zero pairwise differences diverge.

A possible complication with computation of weights (\ref
{eqnadaweights}) is that maximum likelihood estimates
$\hat\mu_i^{\mathrm{(mle)}}$ diverge when team $i$ wins or losses all
its matches. Hence, we suggest to slightly modify
$\hat\mu_i^{\mathrm{(mle)}}$ by adding a small ridge penalty
$\varepsilon
\sum_{i<j} (\mu_i-\mu_j)^2$ to the likelihood (\ref{eqnlik}). In the
applications, we choose a value of~$\varepsilon$ equal to $10^{-4}$.

\subsection{Computation of the ranking lasso solution}\label{sectsolution}

The Lagrangian form of the ranking lasso problem (\ref{eqnrlmin}) is
%
\begin{eqnarray}
\label{eqnlagr} &&(\hat{\fmu}_\lambda, \hat {\tau}_\lambda, \hat{
\ftheta}_\lambda)
\nonumber
\\[-8pt]
\\[-8pt]
\nonumber
&&\qquad=\arg\min \Biggl\{-\ell (\fmu, \tau) +\lambda\sum
_{i<j}^k w_{ij} | \theta_{ij} |
+ \sum_{i<j}^k u_{ij} (
\theta_{ij}-\mu_i+\mu_j ) \Biggr\}.
\end{eqnarray}
The difficulty with the above optimization problem lies
in the computation of the Lagrangian multipliers $u_{ij}$, $i<j$. The
simpler way to overcome this problem is likely the quadratic penalty
method which consists in replacing the Lagrangian term with the
quadratic penalty
%
\begin{equation}
\label{eqnquadratic} \frac{v}{2}\sum_{i<j}^k
( \theta_{ij}-\mu_i+\mu_j )^2,\qquad
v>0.
\end{equation}
The solution of the quadratic penalty method converges
to that of the original problem (\ref{eqnlagr}) as $v$ diverges. However,
numerical analysis literature discourages the use of the quadratic
penalty method, since numerical instabilities may arise for large
values of the penalty coefficient $v$ even if the objective function
is smooth. See, for example, \citeauthor{nocedal2006}
[(\citeyear{nocedal2006}), Section~17.1]. These instabilities motivated the
use of more elaborate methods to solve optimization problems with
linear contrasts, such as the Augmented Lagrangian method introduced by
\citet{hestenes1969} and \citet{powell1969}. We refer the reader to
\citeauthor{nocedal2006} [(\citeyear{nocedal2006}), Section~17.3] for
technical details and further references.

Below we summarize the application of the Augmented Lagrangian method
to the ranking lasso problem. The idea of the Augmented Lagrangian
method is to modify the Lagrangian formulation in a way to add an
explicit estimate of the Lagrangian multipliers $\fu=(u_{12}, \ldots,
u_{k-1 k})$. The target is
achieved by adding the quadratic penalty (\ref{eqnquadratic}) to the
objective function, thus yielding the augmented objective function
%
\begin{eqnarray}
\label{eqnauglagr} F_{\lambda, v}(\fmu, \tau, \ftheta, \fu)&=&-\ell(\fmu, \tau)
+\lambda\sum_{i<j}^k w_{ij} |
\theta_{ij} |
\nonumber
\\[-8pt]
\\[-8pt]
\nonumber
&&{}+ \sum_{i<j}^k u_{ij} (
\theta_{ij}-\mu_i+\mu_j )+\frac{v}{2}
\sum_{i<j}^k ( \theta_{ij}-
\mu_i+\mu_j )^2.
\end{eqnarray}
Hence, differently from the quadratic penalty method, in
the Augmented Lagrangian formulation the quadratic penalty is added to
the Lagrangian term instead of replacing it. Then, the Augmented
Lagrangian method seeks the solution of the original problem through
iteration of the following two steps until convergence:
\begin{enumerate}[\quad]
\item[\textit{minimization step}:] given the current values of the penalty
coefficients $(\fu, v)$, minimize $F_{\lambda, v}(\fmu, \tau,
\ftheta,
\fu)$ with respect to the model parameters $(\fmu, \tau, \ftheta);$
\item[\textit{update step}:] given the current values of the model parameters
$(\fmu, \tau, \ftheta)$, update the Lagrangian multipliers $\fu$ and
the quadratic penalty coefficient~$v$.
\end{enumerate}
The key result of the Augmented Lagrangian method is
that the convergence to the global solution of the original problem
can be assured without increasing $v$ indefinitely if the sequence of
Lagrangian multipliers converges; see Theorem 17.6 of
\citet{nocedal2006}. Thanks to this property, the Augmented Lagrangian
method is a substantially more stable algorithm than the quadratic
penalty method.

Similarly to the illustration by \citet{lian2010} and \citet{ye2011}
about the standard fused lasso problem, the minimization step can be
conveniently performed through cycling between minimization with
respect to the model parameters $(\fmu, \tau)$ for given $\ftheta$ and
minimization of $\ftheta$ for given $(\fmu, \tau)$. Both
these sub-minimization problems have simple and attractive forms. The
first sub-problem is equivalent to computing maximum likelihood
estimates of the Bradley--Terry model with a quadratic penalty
\[
(\hat{\fmu}_\lambda, \hat{\tau}_\lambda)=\arg\min \Biggl\{ -\ell (
\fmu, \tau)+\frac{v}{2}\sum_{i<j}^k (
\hat{\theta}_\lambda-\mu_i+\mu_j )^2
\Biggr\}.
\]
This is a smooth optimization problem that can be handled with
standard numerical algorithms. For the Bradley--Terry model, the
minimization can be efficiently handled by iterated reweighted least
squares.

The second sub-problem consists in solving an ordinary lasso problem
with an orthogonal design. Hence, its solution is computed by the
soft-thresholding operator [\citet{donoho1995}]
\[
\hat{\theta}_{\lambda, ij}=\operatorname{sign}(\tilde{\theta }_{\lambda,
ij})
\biggl( | \tilde{\theta}_{\lambda, ij} | - \frac{\lambda w_{ij}}{v}
\biggr)_{+},\qquad 1<i<j<k,
\]
where $\tilde\theta_{\lambda, ij}=\hat\mu_{\lambda,
i}-\hat\mu_{\lambda, j}- u_{\lambda, ij}/v$ and $(x)_{+}$ indicates
the positive part of $x.$

The second step updates the Lagrangian multipliers and the quadratic
penalty coefficient. The Augmented Lagrangian method provides a simple
recursion for updating the Lagrangian multipliers
\[
u_{\lambda, ij}^{\mathrm{(new)}}=u_{\lambda,
ij}^{\mathrm{(old)}}+v (\hat
\theta_{\lambda, ij}-\hat\mu_{\lambda,
i}+\hat\mu_{\lambda, j} ), \qquad 1<i<j<k.
\]
Finally, the quadratic penalty coefficient $v$ is set equal to the
maximum of the squared $u_{\lambda, ij}^{\mathrm{(new)}},$ so that the
proportion between the two penalty components is preserved. Our
experiments suggest that this simple rule provides stable results.

\subsection{Selection of the lasso penalty}\label{sectselection} The
Augmented Lagrangian method provides a feasible method to compute
estimates of the model parameters $\fmu_\lambda$ and $\tau_\lambda$
and the penalties of $\fu_\lambda$ and $v_\lambda$ for a given value
of the smoothing lasso parameter~$\lambda$. This procedure must be
repeated for a sequence of values of $\lambda$ either in decreasing or
increasing order. An efficient implementation uses the solutions for
a given $\lambda$ as warm starts for minimization of $F_{\lambda,
v}(\fmu, \tau, \ftheta, \fu)$ at a smaller (or larger) value of~$\lambda$.
The remaining task is the selection of a value of
$\lambda,$ which could be somehow optimal in terms of prediction
quality. A viable approach is to base selection of $\lambda$ on the
minimization of an information criterion, such as the Akaike
information criterion,
\[
\operatorname{AIC}(\lambda)=-2 \ell(\hat{\fmu}_\lambda, \hat {\tau
}_\lambda)+ 2 \,\mathrm{df}(\lambda),
\]
or the Schwarz information criterion,
\[
\operatorname{BIC}(\lambda)=-2 \ell(\hat{\fmu}_\lambda, \hat {\tau
}_\lambda)+ \log n \,\mathrm{df}(\lambda).
\]
The effective degrees of freedom $\mathrm{df}(\lambda)$ are estimated as
the number of distinct groups formed with a certain $\lambda,$ by this
way following the previously cited papers on generalized fused lasso
[\citet{gertheiss2010,she2010,tibshirani2010}].

\subsection{Hybrid ranking lasso}\label{sectrefit} Efron et~al. (\citeyear{efron2004})
and \citet{candes2007} suggest hybrid lasso
procedures where sparse methods are used for model selection and then
the selected model is refitted by ordinary least squares. The
refitting procedures are proposed in order to reduce the bias due to
the penalization. \citet{gertheiss2010} advocate refitting for their
generalized fused lasso approach for sparse modeling of categorical
covariates. Following this suggestion, we consider a \textit{hybrid
ranking lasso} method where ranking lasso is used only for groups
selection, while the abilities of the teams are computed by maximum
likelihood constrained so that the abilities of teams in the same group must
be identical. This hybrid procedure is also useful for model
selection. In fact, we found that the computation of the information
criteria such as AIC and BIC at the hybrid ranking lasso estimates
provides more reliable identification of the number of groups. This
finding agrees with \citet{chen2008}, who also suggest to compute
their extended BIC at the hybrid lasso estimates.

\subsection{Uncertainty quantification}\label{sectuncertainty}

One of the advantages of the Bradley--Terry model with respect to a
nonstatistical alternative is the evaluation of the uncertainty about
the difference of estimated abilities of two teams or about the
probability that one team defeats another. If the Bradley--Terry model
is fitted by maximum likelihood, then uncertainty can be quantified by
standard large sample theory through the inverse of the Fisher information.
The quantification of the uncertainty of lasso estimators is more
difficult and, indeed, it is still an open research problem.
Recently, \citet{chatterjee2011} derived a modified version of the
residual bootstrap to approximate the distribution of lasso estimators in
linear regression models. Furthermore, \citeauthor{chatterjee2011}
demonstrated that no modification of the
residual bootstrap is needed for the adaptive lasso method because of
its consistency. Similarly, the parametric bootstrap by
resampling from the Bradley--Terry model with the unknown parameters
replaced by their estimates can be employed for the
evaluation of the uncertainty of adaptive ranking lasso estimators.
Since adaptive ranking lasso estimators are by construction biased, it
is advisable to adjust the bootstrap confidence intervals for bias
[\citet{efron1987}]. The performance of bootstrap confidence intervals
is illustrated in the next section.

\section{NFL 2010--2011 regular season}\label{sectNFL}

\begin{figure}[b]

\includegraphics{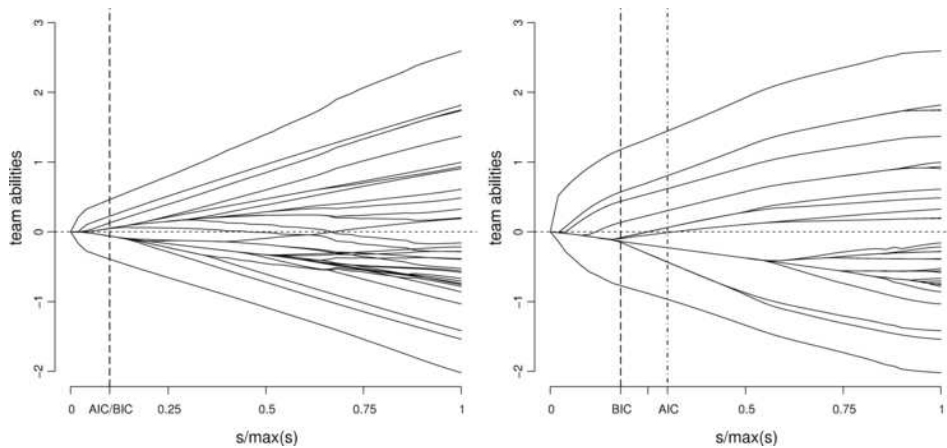}

\caption{NFL regular season 2010--2011. Path plots for the nonadaptive
(left panel) and the adaptive (right panel) ranking lasso. The path is
described in terms of relative bound s$/$max(s), where max(s) is
the minimum value of s such that the ranking lasso solution is
indistinguishable from the unpenalized maximum likelihood
solution. The AIC selection corresponds to the vertical dot-dashed
line, while the BIC selection to the vertical dashed line. For
the nonadaptive case, the two selections coincide.}\label{figpath}
\end{figure}

We start the analysis of the NFL data with the nonadaptive version of
the ranking lasso
where all weights are identical, $w_{ij}=1$ for any $i<j$. The left
panel in Figure~\ref{figpath} shows the path plot of the nonadaptive
estimates for an
increasing sequence of values of the bound $s$. The path is quite
irregular with many crossings between teams and groups along the
clustering process. It would be preferable, instead, to have a smoother
clustering process with intermediate groups formed for relative large
values of $s$ and then fused together in larger
groups when the penalty increases, that is, when $s$ decreases.
%

%
\begin{figure}

\includegraphics{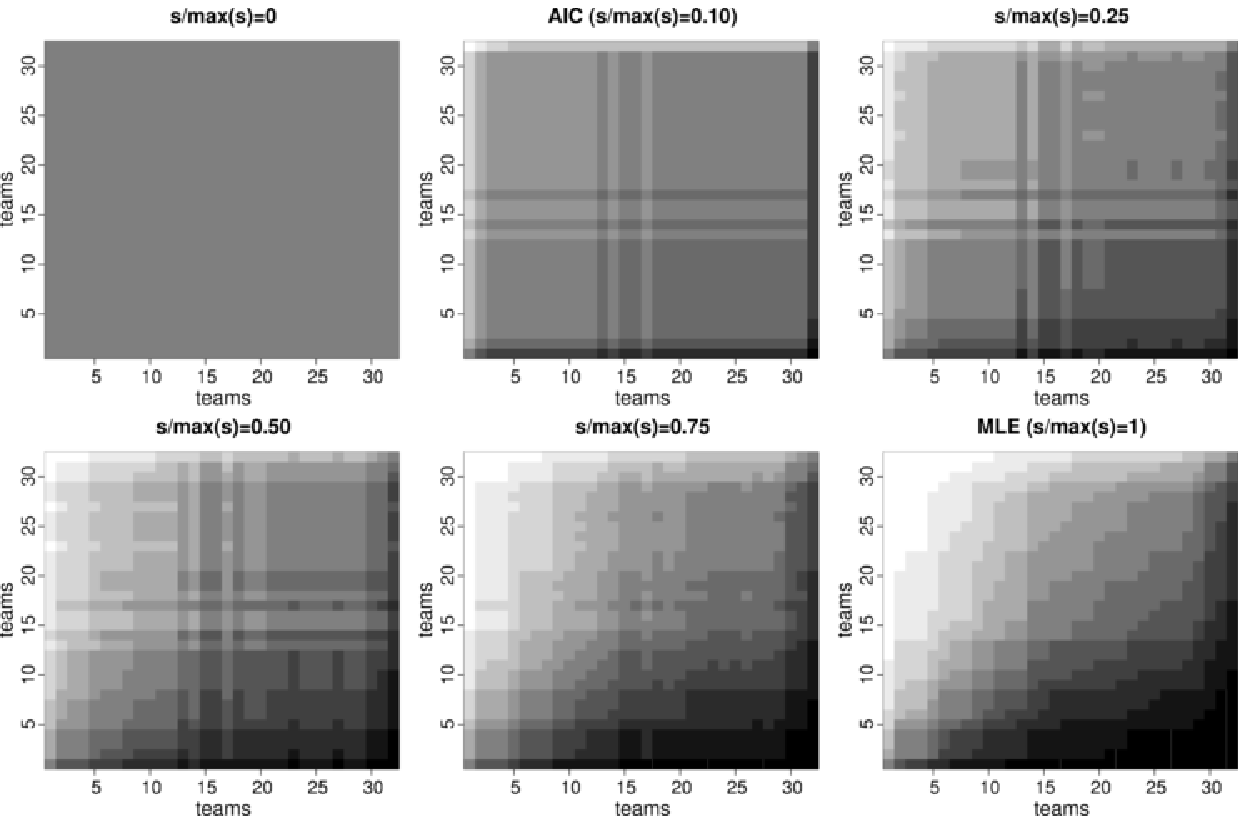}

\caption{NFL regular season 2010--2011. Image plots illustrating
several stages of the nonadaptive ranking lasso path. From top-left to
bottom-right, images correspond to decreasing level of grouping, that
is, increasing values of the
relative bound s$/$max(s). The rows and the columns correspond to the
teams sorted in decreasing order of their maximum likelihood estimated
ability. The pixel of position (row${}={}$r, column${}={}$c) corresponds to the
probability that team r
wins against team c in a match played on a neutral field. Darker
pixels correspond to higher probabilities of victory for the teams on
the row.}\label{figpathnoadaptimage}
\end{figure}

\begin{figure}

\includegraphics{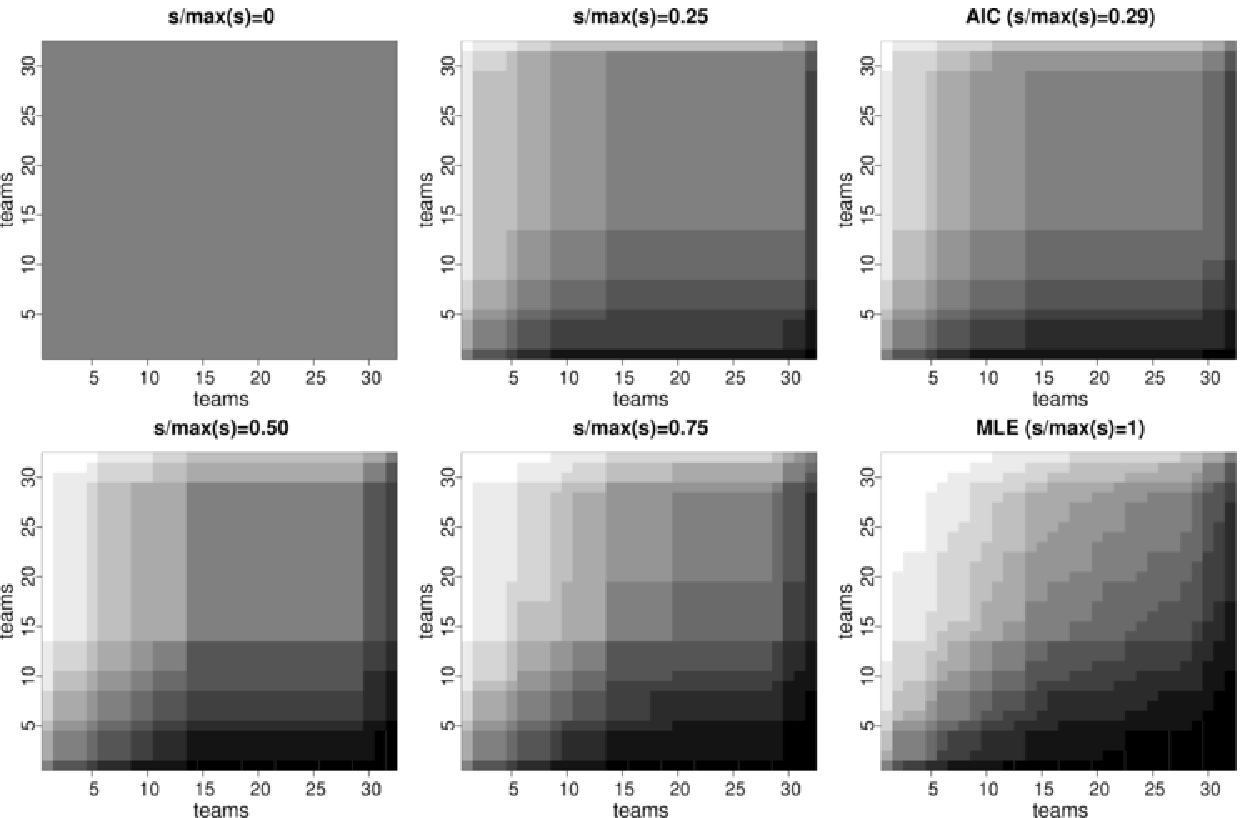}

\caption{NFL regular season 2010--2011. Image plots illustrating
several stages of the adaptive ranking lasso path. From top-left to
bottom-right, images correspond to decreasing level of grouping, that
is, increasing values of the relative bound s$/$max(s). The rows and the
columns correspond to the teams sorted in decreasing order of their
maximum likelihood estimated
ability. The pixel of position (row${}={}$r, column${}={}$c) corresponds to the
probability that team r
wins against team c in a match played on a neutral field. Darker
pixels correspond to higher probabilities of victory for the teams on
the row.}\label{figpathimage}\vspace*{-3pt}
\end{figure}

These drawbacks are fixed by relying on the adaptive version designed
to protect ``true'' large differences between abilities; see the path
displayed in the right panel of Figure
\ref{figpath}.

A useful visualization of the differences between the nonadaptive and
the adaptive solutions is given in Figures
\ref{figpathnoadaptimage} and~\ref{figpathimage}. The image plots
are constructed as follows. The rows and the columns correspond to the
teams sorted in decreasing order of their maximum likelihood estimated
ability, as in Table~\ref{tabranking}. For each image, the pixel of
position $(r,c)$ is the probability that the team in row $r$ beats the
team in column $c$ in a match played on a neutral field (no home
effect),
\[
\pr(Y_{rc}=1)=\frac{\exp(\hat\mu_{\lambda, r}-\hat\mu_{\lambda,
c})}{1+\exp(\hat\mu_{\lambda, r}-\hat\mu_{\lambda, c})},\qquad r,c=1, \ldots, k.
\]
Hence, the diagonal of the image is constant and equal to
$0.5$. Higher values of the probabilities $\pr(Y_{rc}=1)$ correspond
to colors shading off into dark. Figure~\ref{figpathnoadaptimage}
reports the image plots for several stages of the nonadaptive ranking
lasso path, from the complete shrinkage ($s=0$) with all teams
classified into the same group and thus the probability of victory in
any match is $0.5$, the toss of a coin, to the maximum likelihood fit. The
corresponding image plots for the adaptive fit are shown in Figure
\ref{figpathimage}.

The comparison of the image plots in the two figures
provides a clear illustration of the differences of the clustering
process when adaptive weights are employed. Groups formed by the
adaptive ranking lasso (Figure~\ref{figpathimage}) are visualized by
smooth blocks formed by spatially contiguous pixels, thus preserving
the maximum likelihood ranking order. This is a consequence of the
consistency of the adaptive lasso estimation method which assures that,
for a sufficiently large tournament, the sign of the difference between
maximum likelihood and adaptive ranking estimated abilities is the same.
Vice versa, the image plots of the nonadaptive ranking lasso (Figure
\ref{figpathnoadaptimage}) have several spots because teams are
frequently classified in different groups with respect to their closer
neighbors.

\begin{table}
\caption{NFL regular season 2010--2011. Estimated probabilities of
victory for the home team with corresponding 90\% bias corrected
percentile bootstrap confidence intervals for matches the Atlanta
Falcons vs the Baltimore Ravens (home) and the Kansas City Chiefs vs the New England Patriots (home)}\label{tabbootci}
\begin{tabular*}{\textwidth}{@{\extracolsep{\fill}}lcc@{}}
\hline
&\multicolumn{1}{c}{\textbf{Atlanta vs Baltimore}} &\multicolumn{1}{c@{}}{\textbf{Kansas~City vs New~England}}\\
\multicolumn{1}{@{}l}{\textbf{Method}} & \textbf{est. (90\%~c.i.)} & \multicolumn{1}{c@{}}{\textbf{est. (90\%~c.i.)}} \\
\hline
MLE & 0.56 (0.09, 0.90) & 0.96 (0.75, 0.99) \\
AIC & 0.56 (0.32, 0.81) & 0.87 (0.76, 1.00) \\
BIC & 0.56 (0.50, 0.84) & 0.82 (0.78, 1.00) \\
Hybrid AIC & 0.58 (0.15, 0.91) & 0.97 (0.79, 0.99) \\
Hybrid BIC & 0.58 (0.13, 0.93) & 0.97 (0.83, 1.00) \\
\hline
\end{tabular*}
\end{table}

We now move to the interpretation of the adaptive solution reported in
Table~\ref{tabranking}, columns 4--7. AIC selects $9$ groups, while,
as expected, BIC supports a sparser solution with $7$ groups. Both
criteria agree in placing the New England Patriots on a single-team top
group, followed by a group formed by the Atlanta Falcons, the Baltimore Ravens
and the Pittsburgh Steelers. Differences between the two criteria regard
the middle and the bottom part of the ranking. For example, AIC
suggests that the Tampa Bay Buccaneers and the Philadelphia Eagles do better
than the New York Giants, while BIC places these three teams in the
same group together with the Indianapolis Colts and the Miami
Dolphins. Clearly, the abilities estimated by the adaptive ranking
lasso (columns 4--5 in Table~\ref{tabranking}) are considerably
shrunken toward zero with respect to the maximum likelihood
estimates. On the other hand, the hybrid adaptive ranking lasso method
(columns 6--7 in Table~\ref{tabranking}) individuates the same groups
but with estimated abilities that have the same extent of the maximum
likelihood estimates.

The uncertainty of the estimated abilities is evaluated by parametric
bootstrap with $1000$ replications. Table~\ref{tabbootci} reports the
estimated probability of victory of the home teams in the matches
Atlanta vs Baltimore played at Baltimore and Kansas City vs New England played at the
home of the New England Patriots. These two matches were chosen to illustrate the
behavior of the various estimation methods when the match involves
teams with similar ability---Atlanta and Baltimore---or with large
difference---Kansas City and New England. Similar results were
obtained for the other matches.

%
\begin{figure}

\includegraphics{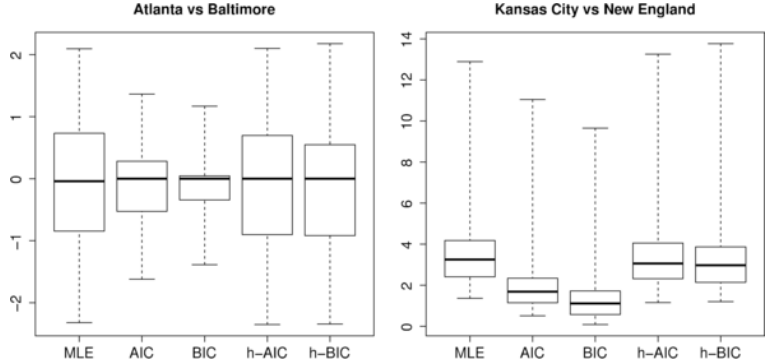}

\caption{NFL regular season 2010--2011. Left panel: boxplots of the
sample distributions of the estimated difference of abilities for the
match between the Atlanta Falcons and the Baltimore Ravens (home) when
the true model parameters correspond to the maximum likelihood
estimates. The boxplots correspond to estimates by maximum likelihood
(\texttt{MLE}), adaptive ranking lasso with \texttt{AIC}
and \texttt{BIC} selection and hybrid adaptive ranking lasso/maximum
likelihood with AIC (\texttt{h-AIC}) and BIC
(\texttt{h-BIC}) selection. Right panel: boxplots of the sample
distributions of the estimated difference of abilities for the match
between the Kansas City Chiefs and the New England Patriots (home).
}\label{figuncertainty}
\end{figure}

We start the discussion from the match between the Atlanta Falcons and
the Baltimore Ravens. The maximum likelihood estimated probability that
the Baltimore Ravens win is $\exp( 1.75-1.82+0.32 )/\{
1+\exp
( 1.75-1.82+0.32 )\}=0.56$. The adaptive ranking lasso method with
either AIC and BIC selection attributes the same ability to the two
teams. The 90\% confidence interval for the victory of Baltimore based on
maximum likelihood is very wide, being equal to $(0.09, 0.90)$.
Adaptive lasso bias-corrected percentile bootstrap confidence intervals
are much shorter: $(0.32, 0.81)$ with AIC selection and $(0.50, 0.84)$
with BIC selection. Instead, hybrid adaptive ranking lasso confidence
intervals are only slightly shorter than the maximum likelihood
confidence interval.

In order to provide insights into the lengths of these confidence
intervals, we estimated the sample distribution of the difference of
the estimated abilities of Atlanta and Baltimore according to the
various estimation methods assuming that the maximum likelihood
estimate is the true model parameter. The above sample distributions
are estimated with $1000$ Monte Carlo simulations and summarized by
the boxplots in the left panel of Figure~\ref{figuncertainty}. Since
the difference of the maximum likelihood estimated abilities for the
two teams is close to zero, $1.75-1.82=-0.07$, the adaptive ranking
lasso estimators have small biases with either AIC and BIC selection.
Furthermore, the shrinkage effect yields a significant reduction in the
variability of the adaptive ranking lasso estimators with respect to
maximum likelihood estimators, as shown by the much smaller height of
the boxes in Figure~\ref{figuncertainty}. Instead, the distributions of
the hybrid adaptive ranking lasso estimators are very similar to the
distribution of the maximum likelihood estimators.

The second match is played by the Kansas City Chiefs on the home
field of the New England Patriots. The difference between the
abilities of these two teams is very large. Indeed, the maximum
likelihood estimate of the probability of New England victory is
$0.96$. The adaptive ranking lasso is somehow more cautious with an
estimated probability of wins equal to $0.87$ and $0.82$ according to
AIC and BIC selections, respectively. The hybrid adaptive ranking lasso
gives estimated probability of victory for New England that is
essentially identical to maximum likelihood. However, the interesting
aspect is that, despite the difference between the estimated
probabilities of victory with maximum likelihood and adaptive ranking
lasso, the bias-corrected confidence intervals are almost identical.

Again, insights into these confidence intervals come from the sample
distribution of the differences of the estimated abilities
assuming\vadjust{\goodbreak}
that the maximum likelihood estimate corresponds to the true model
parameter. Boxplots reported in the right panel of Figure \ref
{figuncertainty} show that in this case adaptive ranking lasso
estimators are significantly biased toward zero with respect to maximum
likelihood and hybrid adaptive ranking lasso estimators. More
interestingly, the height of the boxes of all the five different
estimators is rather similar. Accordingly, \textit{bias-corrected}
bootstrap percentile confidence intervals for the adaptive ranking
lasso estimators are quite similar to those based on maximum likelihood
and hybrid adaptive ranking lasso.\looseness=1

The overall conclusion is that if we compare two teams with close
ability, then the shrinkage of the adaptive ranking lasso provides a
sensible reduction in variability and thus shorter confidence intervals
for the result of the match. Vice versa, if the ability of two teams is
sensibly different, then the adaptive weights allow to obtain
confidence intervals that are essentially equivalent to those obtained
by maximum likelihood. Furthermore, the hybrid method does not seem
particularly convenient in this context because it resembles maximum
likelihood too closely.

These conclusions are coherent with the predictive performance of the
various estimators discussed in the next section.

\subsection{Predictive performance}\label{sectpredict}

We compare the predictive performance of the hybrid and the nonhybrid adaptive
ranking lasso by repeating the following cross-validation exercise
$100$ times:
\begin{longlist}
\item[(1)] form the training set by random sampling without replacing half
the matches in the season;
\item[(2)] determine the estimates of model parameters using the matches in
the training set;
\item[(3)] compute a predictive fit statistic summed over the remaining
matches (the validation set).
\end{longlist}

\begin{figure}

\includegraphics{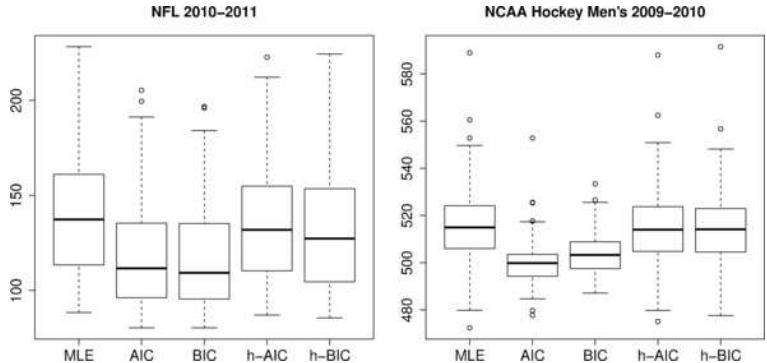}

\caption{Boxplots of the cross-validated negative log-likelihoods
computed at maximum
likelihood (\texttt{MLE}) estimates, adaptive ranking lasso estimates
with \texttt{AIC}
and \texttt{BIC} selection and hybrid adaptive ranking lasso/maximum
likelihood estimates with AIC (\texttt{h-AIC}) and BIC
(\texttt{h-BIC}) selection. Left panel corresponds to the NFL regular
season 2010--2011, right panel to the NCAA College Hockey Men's Division
I 2009--2010.}\label{figlliks}
\end{figure}

As a summary of the forecast's quality in each cross-validation, we
consider the negative of the log-likelihood computed with the matches
in the validation set and model parameters estimated with the matches
in the training set. This choice is, up to a constant term, equivalent
to the Kullback--Leibler divergence and thus consistent with information
model selection criteria.

The left panel of Figure~\ref{figlliks} displays the boxplots of the
$100$ negative log-likelihoods (one for each replication of the
cross-validation experiment)
and Table~\ref{tablliks} provides
some summaries. As clearly shown by boxplots, the shrinkage of adaptive
ranking lasso estimates provides a sensible improvement of the
prediction quality with respect to maximum likelihood predictions.
Summaries in Table~\ref{tablliks} show that the AIC selection improves
on maximum likelihood of about $15\%$ in mean and $19\%$ in median,
while BIC selection does slightly better with an improvement of $16\%$
in mean and $20\%$ in median. Predictions based on the hybrid ranking
lasso, instead, are comparable to those based on maximum likelihood
estimates with a very limited improvement. In summary, this prediction
exercise supports adaptive ranking lasso without refitting because the
method is able to create groups and at the same time increase the
quality of predictions.

Finally, we observe that the percentage of matches which are predicted
better than coin tossing is about $60\%$ for all methods.

\begin{table}[b]
\caption{NFL regular season 2010--2011. Means and medians of the
cross-validated negative log-likelihoods and percentage of predictions
of the correct result which are
better than coin tossing ($\succ\mathrm{coin}$)}\label{tablliks}
\begin{tabular*}{\textwidth}{@{\extracolsep{\fill}}ld{3.2}d{3.2}d{3.2}d{3.2}d{3.2}@{}}
\hline
&& \multicolumn{2}{c}{\textbf{Lasso}} &\multicolumn{2}{c@{}}{\textbf{Hybrid}}
\\[-6pt]
&& \multicolumn{2}{c}{\hrulefill} &\multicolumn{2}{c@{}}{\hrulefill} \\
& \multicolumn{1}{c}{\textbf{MLE}} & \multicolumn{1}{c}{\textbf{AIC}} & \multicolumn{1}{c}{\textbf{BIC}} &
\multicolumn{1}{c}{\textbf{AIC}} & \multicolumn{1}{c@{}}{\textbf{BIC}} \\
\hline
Mean & 139.90 & 119.10 & 117.20 & 135.20 & 131.60
\\
Median & 137.30 & 111.70 & 109.30 & 131.90 & 127.30 \\
$\succ$ coin & 0.59 & 0.60 & 0.58 & 0.60 & 0.60 \\
\hline
\end{tabular*}
\end{table}

\section{Handling ties}\label{sectties} We focused on
the analysis of sports not allowing ties or with no ties observed, as
in the NFL example. However, there are a number of sports where ties are
allowed and occur with a certain frequency. The ranking lasso analysis
of these
sport tournaments follows the lines outlined along the paper,
with the only difference that we need to modify the Bradley--Terry
model so as to handle ties. The match outcome becomes a three-level
ordinal variable that can be arbitrarily coded as
\[
Y_{ijr}=\cases{2, &\quad $\mbox{if team $i$ defeats team $j$,}$\vspace
*{2pt}
\cr
1, &\quad $\mbox{if teams $i$ and $j$ tied,}$\vspace*{2pt}
\cr
0, &\quad $\mbox{if
team $i$ is defeated by team $j$.}$}
\]
Matches with ties can be modeled by some ordinal-valued extension of
the Bradley--Terry model. For example, one may consider a cumulative
link Bradley--Terry model [\citet{agresti2010}]
\[
\pr(Y_{ijr}\leq y_{ijr})=\frac{\exp (\delta_{y_{ijr}}+
h_{ijr}\tau
+ \mu_i -\mu_j  )}{1+\exp (\delta_{y_{ijr}}+ h_{ijr}\tau+
\mu_i -\mu_j  )},
\]
where $-\infty\leq\delta_{0}\leq\delta_{1}\leq\delta_{2}\equiv
+\infty$ are cutpoint parameters, while the other quantities are
defined as in the previous sections of this paper.

Model identifiability now requires an additional contrast. Indeed, for
every match played on a neutral field we must ensure that the
probability that team $i$ defeats team $j$ is equal to the probability
that team $j$ is defeated by team $i$. This condition is guaranteed
when $\delta_{0}=-\delta_{1}$. If no tie is observed, or if it is not
allowed by sport rules, then categories $1$ and $2$ are collapsed,
$\delta_{0}=\delta_{1}=0,$ and the model reduces to the standard
Bradley--Terry model.

\subsection{NCAA college hockey men's division I 2009--2010}\label{secthockey}

We employ the ranking lasso with ties to the regular season
of the NCAA College Hockey Men's Division I 2009--2010. This tournament
comprises $58$ teams partitioned in six conferences, namely, the
Central Collegiate Hockey Association, the Western Collegiate Hockey
Association, the Hockey East, the College Hockey America, the ECAC
Hockey and the Atlantic Hockey. The composite schedule includes within
and between conference games. The total number of matches is $1083$.
The tournament design is highly incomplete. Indeed, about $73.3\%$ of
the $(58 \cdot57)/2=1653$ possible matches are not played, $6.8\%$
are played just once, $10.5\%$ twice and the remaining $9.4\%$ are
played three or more times, with seven matches ($0.4\%$) repeated even
seven times. The tournament is also unbalanced with the total number of
matches per team varying from $31$ to $43$.

Hockey matches may end with ties and these occur with a nonnegligible
frequency. In the regular season of the NCAA College Hockey Men's
Division~I 2009--2010 there were $125$ ties out of the $1083$ matches,
that is, $11.5\%$ of the matches. The home effect also seems quite
relevant because $54.8\%$ of the matches were won by the home team,
$11.6\%$ ended with a tie and $33.5\%$ were won by the visitors. These
numbers do not count the $69$ matches played on a neutral field.

At the end of the season, sixteen teams are qualified for the four
regional semifinals. Hence, the four regional champions compete in the
Frozen Four for the national championship.\vadjust{\goodbreak} The matches' results are
available in the data frame \texttt{icehockey} through the \texttt{R}
package \texttt{BradleyTerry2} [\citet{turner2011}]. As reported in the
help pages of this package, the NCAA selection system has been the
source of several criticisms because there is no agreement that it
correctly accounts for the highly irregular design of the tournament.
The ranking based on the Bradley--Terry model is seen as a sensible
alternative to the NCAA selection mechanism.

The maximum likelihood estimates of the home field parameter $\tau$ and
the threshold parameter $\delta_1$ are both strongly significant:
$\hat
\tau^{(\mathrm{mle})}=0.402$ with a standard error of $0.066$ and
$\hat
\delta_1^{(\mathrm{mle})}=0.288$ with a standard error of $0.024$. The
maximum likelihood estimates of the teams abilities, under the sum
contrast, are listed in the third column of Table \ref
{tabrankinghockey}. According to the maximum likelihood fit of the
Bradley--Terry model with ties, the best team is Denver, followed by
Miami (Ohio), Wisconsin and Boston College. The last two teams were the
finalists of the national championship won by Boston College on April
4, 2010.

\begin{table}
\caption{American College Hockey Men's Division I composite schedule
2009--2010. For each team, the table
displays the record and the ability estimated by maximum likelihood
(\texttt{MLE}), by adaptive ranking lasso (\texttt{lasso}) and by
hybrid adaptive ranking lasso/maximum likelihood (\texttt{hybrid}).
Results are
shown with both \texttt{AIC} and \texttt{BIC} model selection. Teams
qualified for NCAA Regional semifinals are marked by symbol $^\dag$}\label{tabrankinghockey}
%
\begin{tabular*}{\textwidth}{@{\extracolsep{\fill}}lcd{2.2}d{2.2}d{2.2}d{2.2}d{2.2}@{}}
\hline
&&& \multicolumn{2}{c}{\textbf{Lasso}} & \multicolumn{2}{c@{}}{\textbf{Hybrid}}
\\[-6pt]
&&& \multicolumn{2}{c}{\hrulefill} & \multicolumn{2}{c@{}}{\hrulefill} \\
\multicolumn{1}{@{}l}{\textbf{Teams}} &
\multicolumn{1}{c}{\textbf{Record}} & \multicolumn{1}{c}{\textbf{MLE}} & \multicolumn{1}{c}{\textbf{AIC}} & \multicolumn{1}{c}{\textbf{BIC}} &
\multicolumn{1}{c}{\textbf{AIC}} & \multicolumn{1}{c@{}}{\textbf{BIC}} \\
\hline
Denver$^\dag$ & 27--4--9 & 1.65 & 0.58 & 0.41 & 1.38 & 1.36 \\
Miami (Ohio)$^\dag$ & 27--7--7 & 1.60 & 0.58 & 0.41 & 1.38 & 1.36 \\
Wisconsin$^\dag$ & 25--4--10 & 1.53 & 0.58 & 0.41 & 1.38 & 1.36 \\
Boston College$^\dag$ & 25--3--10 & 1.43 & 0.58 & 0.41 & 1.38 & 1.36
\\
North Dakota$^\dag$ & 25--5--12 & 1.37 & 0.58 & 0.41 & 1.38 & 1.36 \\
St Cloud State$^\dag$ & 23--5--13 & 1.10 & 0.18 & 0.09 & 0.60 & 0.56
\\
New Hampshire$^\dag$ & 17--7--13 & 0.89 & 0.18 & 0.09 & 0.60 & 0.56
\\
Minnesota Duluth & 22--1--17 & 0.87 & 0.18 & 0.09 & 0.60 & 0.56 \\
Bemidji State$^\dag$ & 23--4--9 & 0.87 & 0.18 & 0.09 & 0.60 & 0.56 \\
Michigan$^\dag$ & 25--1--17 & 0.86 & 0.18 & 0.09 & 0.60 & 0.56 \\
Colorado College & 19--3--17 & 0.86 & 0.18 & 0.09 & 0.60 & 0.56 \\
Northern Michigan$^\dag$ & 20--8--12 & 0.81 & 0.18 & 0.09 & 0.60 &
0.56 \\
Vermont$^\dag$ & 17--7--14 & 0.79 & 0.18 & 0.09 & 0.60 & 0.56 \\
Ferris State & 21--6--13 & 0.77 & 0.18 & 0.09 & 0.60 & 0.56 \\
Minnesota & 18--2--19 & 0.74 & 0.18 & 0.09 & 0.60 & 0.56 \\
Alaska$^\dag$ & 18--9--11 & 0.74 & 0.18 & 0.09 & 0.60 & 0.56 \\
Cornell$^\dag$ & 21--4--8 & 0.73 & 0.18 & 0.09 & 0.60 & 0.56 \\
Maine & 19--3--17 & 0.66 & 0.18 & 0.09 & 0.60 & 0.56 \\
UMass-Lowell & 19--4--16 & 0.64 & 0.18 & 0.09 & 0.60 & 0.56 \\
Yale$^\dag$ & 20--3--9 & 0.60 & 0.18 & 0.09 & 0.60 & 0.56 \\
Michigan State & 19--6--13 & 0.58 & 0.18 & 0.09 & 0.60 & 0.56 \\
Boston University & 18--3--17 & 0.57 & 0.18 & 0.09 & 0.60 & 0.56 \\
Nebraska-Omaha & 20--6--16 & 0.57 & 0.18 & 0.09 & 0.60 & 0.56 \\
Massachusetts & 18--0--18 & 0.56 & 0.18 & 0.09 & 0.60 & 0.56 \\
Northeastern & 16--2--16 & 0.51 & 0.18 & 0.09 & 0.60 & 0.56 \\
Ohio State & 15--6--18 & 0.45 & 0.18 & 0.09 & 0.60 & 0.56 \\
Minnesota State & 16--3--20 & 0.43 & 0.18 & 0.09 & 0.60 & 0.56 \\
Merrimack & 16--2--19 & 0.40 & 0.18 & 0.09 & 0.60 & 0.56 \\
Union (New York) & 21--6--12 & 0.29 & 0.18 & 0.09 & 0.60 & 0.56 \\
Notre Dame & 13--8--17 & 0.16 & 0.10 & 0.09 & 0.10 & 0.56 \\
Lake Superior & 15--5--18 & 0.15 & 0.10 & 0.09 & 0.10 & 0.56 \\
Alaska Anchorage & 11--2--23 & -0.00 & -0.09 & -0.04 & -0.34 & -0.34
\\
St. Lawrence & 19--7--16 & -0.17 & -0.09 & -0.04 & -0.34 & -0.34 \\
Providence & 10--4--20 & -0.19 & -0.09 & -0.04 & -0.34 & -0.34 \\
Rensselaer & 18--4--17 & -0.20 & -0.09 & -0.04 & -0.34 & -0.34 \\
Quinnipiac & 20--2--18 & -0.24 & -0.09 & -0.04 & -0.34 & -0.34 \\
Western Michigan & 8--8--20 & -0.24 & -0.09 & -0.04 & -0.34 & -0.34
\\
Colgate & 15--6--15 & -0.34 & -0.09 & -0.04 & -0.34 & -0.34 \\
\hline
\end{tabular*}
\end{table}
\setcounter{table}{3}
\begin{table}
\caption{(Continued)}
%
\begin{tabular*}{\textwidth}{@{\extracolsep{\fill}}lcd{2.2}d{2.2}d{2.2}d{2.2}d{2.2}@{}}
\hline
&&& \multicolumn{2}{c}{\textbf{Lasso}} & \multicolumn{2}{c@{}}{\textbf{Hybrid}}
\\[-6pt]
&&& \multicolumn{2}{c}{\hrulefill} & \multicolumn{2}{c@{}}{\hrulefill} \\
\multicolumn{1}{@{}l}{\textbf{Teams}} &
\multicolumn{1}{c}{\textbf{Record}} & \multicolumn{1}{c}{\textbf{MLE}} & \multicolumn{1}{c}{\textbf{AIC}} & \multicolumn{1}{c}{\textbf{BIC}} &
\multicolumn{1}{c}{\textbf{AIC}} & \multicolumn{1}{c@{}}{\textbf{BIC}} \\
\hline
Rochester Institute of Technology$^\dag$ & 26--1--11 & -0.39 & -0.09 &
-0.04 & -0.34 & -0.34 \\
Alabama-Huntsville$^\dag$ & 12--3--17 & -0.49 & -0.09 & -0.04 &
-0.34 & -0.34 \\
Robert Morris & 10--6--19 & -0.50 & -0.09 & -0.04 & -0.34 & -0.34 \\
Niagara & 12--4--20 & -0.51 & -0.09 & -0.04 & -0.34 & -0.34 \\
Princeton & 12--3--16 & -0.56 & -0.09 & -0.04 & -0.34 & -0.34 \\
Brown & 13--4--20 & -0.61 & -0.09 & -0.04 & -0.34 & -0.34 \\
Bowling Green & 5--6--25 & -0.76 & -0.25 & -0.10 & -0.93 & -0.92 \\
Sacred Heart & 21--4--13 & -0.80 & -0.09 & -0.04 & -0.34 & -0.34 \\
Harvard & 9--3--21 & -0.89 & -0.25 & -0.10 & -0.93 & -0.92 \\
Dartmouth & 10--3--19 & -0.89 & -0.25 & -0.10 & -0.93 & -0.92 \\
Michigan Tech & 5--1--30 & -1.03 & -0.42 & -0.24 & -1.30 & -1.30 \\
Clarkson & 9--4--24 & -1.06 & -0.42 & -0.24 & -1.30 & -1.30 \\
Air Force & 16--6--15 & -1.27 & -0.25 & -0.10 & -0.93 & -0.92 \\
Canisius & 17--5--15 & -1.31 & -0.25 & -0.10 & -0.93 & -0.92 \\
Mercyhurst & 15--3--20 & -1.59 & -0.42 & -0.24 & -1.30 & -1.30 \\
Army & 11--7--18 & -1.60 & -0.42 & -0.24 & -1.30 & -1.30 \\
Holy Cross & 12--6--19 & -1.71 & -0.42 & -0.24 & -1.30 & -1.30 \\
Bentley & 12--4--19 & -1.78 & -0.42 & -0.24 & -1.30 & -1.30 \\
Connecticut & 7--3--27 & -2.44 & -1.17 & -0.97 & -2.19 & -2.18 \\
American International & 5--4--24 & -2.60 & -1.17 & -0.97 & -2.19 &
-2.18 \\
\hline
\end{tabular*}
\end{table}

Adaptive ranking lasso estimates of team abilities with or without
refitting are listed in columns from four to seven of Table \ref
{tabrankinghockey}. AIC selects seven groups with a top group formed by
the five teams with higher maximum likelihood estimates, namely,
Denver, Miami, Wisconsin, Boston College and North Dakota. BIC instead
suggests a slightly sparser solution with six groups. The only
difference between AIC and BIC is that the latter rates Lake Superior
and Alaska Anchorage at the same level of the preceding group in the
AIC ranking.

The rankings obtained with the adaptive lasso penalization differ from
the maximum likelihood ranking for a few teams at the bottom of the
ranking. This result is not surprising. Both maximum likelihood and
adaptive ranking lasso yield consistent estimation of teams' abilities
and, thus, they are expected to converge to the same ranking for
\textit{sufficiently large} tournaments, but for a finite tournament
differences between the two rankings may occur. Given the strong
incompleteness of the NCAA hockey tournament, the few observed
differences between the two rankings are reasonable.

The predictive performance of the adaptive ranking lasso with the NCAA
hockey tournament data is evaluated by the same cross-validation
exercise previously used for the NFL example, as described at the
beginning of Section~\ref{sectpredict}. The right panel of Figure \ref
{figlliks} displays the boxplots of the cross-validated negative
log-likelihoods computed at the various estimators. The figure
illustrates the outstanding predictive performance of the adaptive
ranking lasso which largely \mbox{outperforms} predictions based on maximum
likelihood. Differently from the previously analyzed NFL example, the
larger number of teams and the strong irregularity of the tournament
makes more evident the usefulness of the grouping effect induced by the
lasso. Furthermore, the results strongly support the selection of the
lasso penalty by AIC, while in the NFL application, predictions based
on AIC and BIC were essentially of the same quality.

\section{Conclusions}\label{sectconclusions}

Lasso and its many variants have provided successful solutions to
model selection in a variety of high-dimensional problems. In this
paper we suggested a further use of the lasso ideas in the
context of ranking contestants participating in a tournament. We
showed how a generalized fused lasso penalty can be used for enhancing
rankings derived from paired comparison models. The proposed adaptive ranking
lasso method produces ranking in groups in a way that teams with similar
ability are shrunk to the same common level.

Uncertainty of ranking lasso estimates can be evaluated by means of a
parametric bootstrap. Our results support the idea that, as expected,
the lasso-based estimates are more precise with respect to maximum
likelihood estimates, in particular, when the true abilities of two
teams are equal or nearly equal.

Lasso and other shrinkage methods are often motivated by superior
predictive performance with respect to standard maximum likelihood.
This is also the case of the proposed adaptive ranking lasso method. An
empirical study suggests that ranking in groups induced by the adaptive
ranking lasso
produces forecasts of future matches whose quality is sensibly better than
predictions based on maximum likelihood.

Although this paper is addressed to sport
tournaments, we think that the discussed methodology can be of interest in
many other ambits where rankings have to be derived from preference
data. Further, the results in this paper can also be of interest
because they highlight the benefits of adaptive versions of the lasso
method as suggested by \citet{zou2006}.

\section*{Acknowledgments}
The authors thank M. Cattelan and A. Guolo, two anonymous referees, one
Associate Editor and the Editor Susan M. Paddock for helpful comments
and suggestions.

%

%

%


\printaddresses

\end{document}